\documentclass[12pt,unsortedaddress,superscriptaddress,a4paper,nofootinbib,%
nobalancelastpage,preprintnumbers,showpacs,tightenlines,eqsecnum]{revtex4}
\usepackage{amsmath, amsbsy}
\usepackage{bbm}
\usepackage[pdftex]{graphicx}
\def\RE{\mathop{\Re{\rm e}}\nolimits}
\def\vec#1{\mathbf{#1}}
\def\be{\begin{equation}}
\def\bel#1{\begin{equation}\label{#1}}
\def\ee{\end{equation}}
\def\d{\mathrm{d}}
\def\dd{\,\mathrm{d}}
\def\fracd#1#2{\frac{\displaystyle#1}{\displaystyle #2}}
\begin{document}
\title{Low-energy potential scattering
 in two and three dimensions}
\author{N. N. Khuri}
\email{khuri@mail.rockefeller.edu}
\affiliation{Department of Physics,
The Rockefeller University, New York, New York 10021}
\author{Andr\'e Martin}
\email{martina@mail.cern.ch}
\affiliation{TH Division, CERN, CH - 1211 Geneva 23}
\author{J.-M. Richard}
\email{jean-marc.richard@lpsc.in2p3.fr}
\affiliation{Laboratoire de Physique Subatomique et Cosmologie,
 IN2P3-CNRS, Universit\'e Joseph Fourier,  INPG,
 53, avenue des Martyrs, Grenoble, France}
\author{Tai Tsun Wu}
\email{Tai.Tsun.Wu@cern.ch}
\affiliation{Gordon McKay Laboratory, Harvard University,
Cambridge, Massachusetts 02138-2901}
\affiliation{TH Division, CERN, CH - 1211 Geneva 23}

\date{\today}
\begin{abstract}
Conditions are established for the existence of a scattering length and  an effective range in the low-energy expansion of the S-wave phase-shift of a central potential in two and three dimensions.
The behavior of the phase-shift as a function of  the momentum is also derived for longer-range power-law potentials which do not fulfill these conditions.
\end{abstract}
\maketitle
\section{Introduction}
Ten years ago, low-energy scattering in two spatial dimensions (2D) was studied for a large class of potentials \cite{Chadan:1998qm}.
It was found that the S-wave phase-shift as a function of the momentum $k$,  $\delta_0(k)$,  has a universal behavior
\bel{eq:tand1}
\tan\delta=\frac{\pi}{2}\,\frac{1}{\ln k}+ \frac{\mathit{o}(1)}{\ln k}~,
\ee
as $k\to 0$. See, e.g., \cite{0305-4470-19-12-018,PhysRevLett.56.900,0696.35040} for earlier studies, and  \cite{Khuri:2004hi} for the case of non-central potentials.

This behavior is radically different from that for three spatial dimensions (3D), where $\delta_0(k)\sim - a k$, where $a$ is the scattering length.

This universal behavior (\ref{eq:tand1}) is valid for most potentials in 2D, but there is a subclass of potentials where Eq.~(\ref{eq:tand1})  does not hold. Special cases of (\ref{eq:tand1})  have been studied earlier~\cite{Chadan:1998qm}.

This universal behavior is the underlying reason why wire antennas are very efficient and widely used \cite{King:104477}.

It is difficult to obtain an approximate value for $\tan\delta$ from this asymptotic formula (\ref{eq:tand1}). The basic reason is that, since $k$ has the dimension of inverse length, the value of $\ln k$ depends on the length unit used. In order to  get a good approximate value, we need some knowledge of the next to leading term, i.e., we would like to rewrite the above result in the form
\bel{eq:tand2}
\tan\delta\simeq\frac{\pi}{2}\;\frac{1}{\ln(k/k_0)}~,
\ee
for small $k$, and obtain more information about the scale $k_0$. What can $k_0$ be? Clearly it cannot be universal. And it is reasonable to think that the value of $k_0$ is related to a scattering length. 

We recall that  in 3D, the scattering length appears in two different ways. For small $k$, we have $\delta\sim -k a$ (with the  convention that $a>0$ for a repulsive potential). But in addition, the scattering length, $a$, can be defined by the zero of the reduced zero-energy asymptotic wave-function
\bel{eq:phias}
u_\text{asym}=C(r-a)~,
\ee
We use this latter route to define a scattering length in 2D. We shall see that the scale in (\ref{eq:tand2}) is proportional to the inverse of $a$, the value for which the zero-energy asymptotic wave function vanishes. 

In Sec.~\ref{se:finite}, we treat, for pedagogical reasons, the simple case of a potential with finite range. In Sec.~\ref{se:a-2d}, we derive sufficient conditions on the potential for the existence of  the scattering length in 2D. Section~\ref{se:2g} is devoted to the case of weak coupling, where very amazing effects are observed. It also contains numerical illustrations which played a crucial in our preliminary investigations. 

Next, in Sec.~\ref{se:2R}, following the analogy with 3D, we question the validity of an ``effective range''  formula, and find sufficient (and sometimes necessary) conditions for the existence of the effective range. These conditions require a sufficiently-rapid average decrease of the potential at large distance $r$. 
We also find what happens when the potential decreases less rapidly, behaving like negative powers of $r$ at large $r$.
Finally, in Sec.~\ref{se:3R}, we return to the effective range in 3D, for the existence of which we believe to have the best sufficient conditions, in spite of the fact that the notion of effective range is known since a very long time and is well documented \cite{Blatt:628052,Joachain:104904,Mott:104647,1983AnPhy.148..308A,0528.35076}.
\section{The case of a finite range potential}\label{se:finite}
As an example,  we discuss the case of a finite range potential
\bel{eq:frp}
V(r)=0~,\quad\text{for}\quad r>R~,
\ee
The wave function for $r>R$, is given by
\bel{eq:phiext}
\Psi(k,r)=C_0\left[J_0(kr)\cos\delta - Y_0(kr)\sin\delta\right]~,
\ee
which has the asymptotic form
\bel{eq:psiextas}
\Psi \xrightarrow[r\to\infty]{} \,C_0\,\sqrt{\frac{\pi}{2 k r}} \,\cos(k r + \delta -\pi/4)~.
\ee
We need to match at $r=R$ the external wave function (\ref{eq:phiext}) to the internal one, $\Psi(k,r)$,
\bel{eq:match1}
\frac{\partial_r\Psi(k,R)}{\Psi(k,R)}=
\frac{k\left[ J_0'(kR)\sin\delta - Y_0'(kR)\cos\delta\right]}{J_0(kR)\sin\delta - Y_0(kR)\cos \delta}~,
\ee
where $\partial_r$ indicates derivation with respect to the second variable.

By an extension of Poincar\'e's theorem under conditions on $V$ to be specified later, we have, if 
$\Psi(0,R)\neq 0$
\bel{eq:match2}
\frac{\partial_r\Psi(0,R)}{\Psi(0,R)}=\frac{\partial_r\Psi(k,R)}{\Psi(k,R)}+ \mathit{O}(k^2)~.
\ee
On the other hand, for small $z$, the Bessel functions are given by
\bel{eq:Bessel-small}
\begin{split}
J_0(z)&=1+\mathit{O}(z^2)~,\\
Y_0(z)&=\frac{2}{\pi}J_0(z)\left[\ln\left(\frac{z}{2}\right)+\gamma\right]+ \mathit{O}(z^2)~,
\end{split}\ee
where $\gamma$ is  Euler's constant.

The zero-energy wave function for $r>R$ is given by
\bel{eq:z-e-w-f}
\Psi(0,r)\propto\ln \left({r}/{a}\right)~.
\ee
Combining Eqs.(\ref{eq:match1})-(\ref{eq:z-e-w-f}), we obtain
\bel{eq:cotd}
\cot\delta\simeq \frac{2}{\pi}\left[\ln\left(\frac{k a}{2}\right) + \gamma\right]~,\quad\text{as}\quad k\to0~.
\ee

One should note that the 2D scattering length $a$ gives the value of $r$ where the asymptotic zero-energy wave function vanishes, in complete analogy to the 3D case.
A major difference between the 2D 	and 3D cases is that, in 2D, the scattering length $a$ is \emph{always non-negative}.
\section{Existence of the scattering length in 2D}\label{se:a-2d}
In this section, we start with the S-wave ($m=0$) solutions, $u(k,r)$, which satisfy
\bel{eq:radialeq}
\left[\frac{\d^2}{\d r^2}+\frac{1}{4 r^2}+k^2 -g V(r)\right] u(k,r)=0~.
\ee
Of the two independent solutions of (\ref{eq:radialeq}), under conditions on $V(r)$ to be specified below, we choose as a regular solution
\bel{eq:ureg}
u(k,0)=0~,\quad u(k,r)\sim \sqrt{r} \quad\text{as}\quad r\to 0~,
\ee
corresponding to a wave function $\Psi(k,r)=u(k,r)/\sqrt{r}$ which is finite at the origin. 

We introduce  the Green's function $G(r,r')$ for $r,\,r'>0$, defined by
\bel{eq:rGr1}
\left[\frac{\d^2}{\d r^2}+\frac{1}{4 r^2}+k^2\right] G(r,'r)=\delta(r-r')~.
\ee
This $G$ is given by
\bel{eq:rGr2}
G(r,r')=-\frac{\pi}{2}\,\sqrt{r r'} \left[J_0(kr)Y_0(kr') - J_0(kr') Y_0(k r)\right] \Theta(r'-r)~.
\ee
The Volterra integral equation that gives the solution of Eq.~(\ref{eq:radialeq}) guaranteeing Eq.~(\ref{eq:ureg}) is
\bel{eq:Volt1}
u(k,r)=u_0(k,r)+g  \int_0^r\dd r'\,G(r,r')\,V(r')\, u(k,r')~,
 \ee
where we set $u_0(k,r)=\sqrt{r}\,J_0(k r)$. 

We proceed to find a bound on $|G(r,r')|$ which is uniform in $k$. We start with the well-known result 
\bel{eq:R}
Y_0(z)=\frac{2}{\pi}J_0(z)\left[\ln (z/2)+\gamma\right]+R(z)~,
\ee
where $R(z)\to 0$ as $z\to 0$ \cite{Stegun:102084}.
In Appendix \ref{se:Appa}, we prove that for any $z$
\bel{eq:boundR}
|R(z)|<\frac{8}{3\pi}~.
\ee
Substituting Eqs.~(\ref{eq:R}) and (\ref{eq:boundR}) into (\ref{eq:rGr2}), we obtain
\bel{eq:boundG}
|G(r,r')|<\sqrt{r r'}\left[\ln\left(\frac{r}{r'}\right)+3\right]~.
\ee
Now it is easy to show that 
\bel{eq:lnrrp} 
0<\ln\left(\frac{r}{r'}\right)+3\le \ln^+r+\ln^-r'+3\le\left(\sqrt{3}+\frac{\ln^+r}{\sqrt3}\right)\left(\sqrt3 + 
\frac{\ln^-r'}{\sqrt3}\right)~,
\ee
where
\be
\ln^+ r=\begin{cases}\ln r & \text{if $r\ge1$}\\
0&\text{if $r< 1$}\end{cases}
\qquad
\ln^- r=\begin{cases}
0&\text{if $r\ge 1$}\\
-\ln r & \text{if $r<1$}\end{cases}~.
\ee
Hence
\bel{eq:boundG1}
|G(r,r')|<\sqrt{r r'} \left(\sqrt3+\frac{\ln^+r'}{\sqrt3}\right)
                              \left(\sqrt3+\frac{\ln^-r'}{\sqrt3}\right)~.
\ee

Using (\ref{eq:boundG1}), we can deduce from (\ref{eq:Volt1}) an integral inequality
\bel{eq:intin1}
|u(r)|<\sqrt{r} + \sqrt{r}\left(\sqrt3+\frac{\ln^+ r}{\sqrt3}\right) I(r)
\ee
with
\bel{eq:defI}
I(r)=g\int_0^r \left(\sqrt3+\frac{\ln^- r'}{\sqrt3}\right)|V(r')|\,|u(r')|\dd r'~.
\ee
This inequality can be easily integrated, giving
\bel{eq:intin2}
I(r)<\exp\left[g\int_0^r r'\left(3+\ln^-r'\right)|V(r')|\dd r'\right]~,
\ee
and
\bel{eq:intin3}
|u(r)|<2 \sqrt{r} \left(1+\ln^+ r\right)\exp\left[g\int_0^r r'\left(3+|\ln r'|\right)|V(r')|\dd r'\right]~,
\ee
So, if the integral 
\bel{eq:intV1}
\int_0^\infty r\left(3+|\ln r|\right)|V(r)|\dd r~,
\ee
converges, the iterative solution of (\ref{eq:Volt1}) exists (see, for instance, \cite{D'Alfaro:103516}).

We now proceed to derive an expression for the scattering length in 2D. Recalling the remarks in the introduction, the scattering length in 3D is given by the value of $r$ for which the asymptotic reduced wave-function, i.e., $u_\text{asym}=A+B\,r$,  vanishes. In the 2D case, 
\bel{eq:uas2d}
u_\text{asym}=(A+B\,\ln r)\,\sqrt{r}~,
\ee
and we thus define the scattering length as 
\bel{eq:defa2d}
a=\exp\left[-A/B\right]~.
\ee
This shows explicitly that the scattering length is non negative. It is zero or infinity when $B=0$.

The reduced radial equation in a partial wave reads $-u"+ (\alpha/r^2+V(r))u(r)=k^2 u(r)$ with $\alpha \ge -1/4$. In the usual cases, $\alpha>-1/4$, and the sign of the phase shift reflects how the effective potential departs from the pure centrifugal term.  Here,  $\alpha=-1/4$ is the minimal admissible value. This is why, the departure due to $V(r)$ seen in the low-energy phase-shift has always the same sign.

The zero-energy limit of the integral equation (\ref{eq:Volt1}), with $u_0(r)=u(0,r)$, is
\bel{eq:Volt1-0}
u_0(r)=\sqrt{r}+g\int_0^r \sqrt{rr'}\,\ln\left(\frac{r}{r'}\right) V(r')\, u_0(r')\dd r'~,
\ee
It can be rewritten as
\bel{eq:Volt1-0a}
\frac{u_0(r)}{\sqrt{r}}=
1+g\int\limits_0^\infty\sqrt{r'}\,\ln\left(\frac{r}{r'}\right)
V(r')\, u_0(r')\dd r'
-g \int\limits_r^\infty\sqrt{r'}\,\ln\left(\frac{r}{r'}\right) V(r')\, u_0(r')\dd r'~.
\ee
We define $X_1$ and $X_2$ as
\begin{align}
X_1(g)&=\int_0^\infty \sqrt{r} \,V(r)\, u_0(r)\dd r~,\label{eq:defX1}\\
X_2(g)&=\int_0^\infty \sqrt{r} \,\ln r\,\,V(r)\, u_0(r)\dd r~,\label{eq:defX2}
\end{align}
 Because of (\ref{eq:intin3}), valid for any $k$, these integrals converge. For the time being, we shall assume that $X_1(g)$ differs from zero. If $X_1(g)=0$, this constitutes the exceptional case discussed in Ref.~\cite{Chadan:1998qm}.

From (\ref{eq:Volt1-0a}), we now get 
\bel{eq:uzX1X2}
\frac{u_0(r)}{\sqrt{r}}=1-g\,X_2(g)+ g\,X_1(g)\,\ln r + g \int_r^\infty \sqrt{r'}\,\ln\frac{r}{r'}\,V(r')\, u_0(r')\dd r'~.
\ee
Hence, as $r\to\infty$, if we set
\bel{eq:lna}
\ln a=\frac{g\,X_2(g)-1}{g\,X_1(g)}~,
\ee
the asymptotic $u_0(r)$ will have a zero at $r=a$.    For any $|g|>0$, given our bound on $u_0(r)$, $\ln a$ exists and is finite, and given by
\bel{eq:afromlna}
a=\exp\left[\frac{g X_2-1}{g X_1}\right]~.
\ee

The next question  is to see how this length $a$ appears in the low-energy  formula of $\delta(k)$. The asymptotic form of $u(k,r)$  is 
\bel{eq:ukasy}
\lim_{\scriptstyle r\to\infty\atop \scriptstyle k\, \text{fixed}} u(k,r)=u_\text{asy}(k,r)
\propto\sqrt{r}\left[ J_0(kr)\,\cos\delta(k) - Y_0(ky)\,\sin\delta(k)\right]~,
\ee
Hence from the integral equation for $u(k,r)$, 
\bel{eq:cotdelta-1}
\cot\delta(k)=\lim_{r\to\infty}\fracd%
{1-\frac{\pi}{2}\,g \int_0^r Y_0(kr')\,V(r')\,u(k,r')\,\sqrt{r'}\dd r'}
{-\frac{\pi}{2}\,g \int_0^r J_0(kr')\,V(r')\,u(k,r')\,\sqrt{r'}\dd r'}~,
\ee
and from  (\ref{eq:R}) we get
\bel{eq:cotdelta-2}
\cot\delta(k)=\frac{2}{\pi}\left(\ln (k/2)+\gamma\right)+
\lim_{r\to \infty}
\fracd%
{1-g \int_0^r \left[ \ln r' +\frac{\pi}{2}R(kr)\right] V(r )\,u(k,r )\,\sqrt{r}\dd r}
{-\frac{\pi}{2}\,g \int_0^r  J_0(kr)\,V(r)\,u(k,r)\,\sqrt{r}\dd r}~,
\ee
Both integrals in (\ref{eq:cotdelta-2}) exist given our bound on $|u(k,r)|$ in Eq.~(\ref{eq:intin3}) and the fact that $R(kr')|<1$, and if we also impose the second condition on the potential, namely
\bel{eq:intV2}
\int_1^\infty  \ln^2 r \,|V(r)|\,r\dd r<\infty,
\ee
we have
\bel{eq:cotdelta-3}
\cot\delta(k)=\frac{2}{\pi}\left[\ln\left(\frac{k}{2}\right)+\gamma\right]+
\fracd%
{1-g \int_0^\infty \left[ \ln r' +\frac{\pi}{2}R(kr)\right] V(r )\,u(k,r )\,\sqrt{r}\dd r}
{-\frac{\pi}{2}\,g \int_0^\infty  J_0(kr)\,V(r)\,u(k,r)\,\sqrt{r}\dd r}~,
\ee
Next we show that $\lim_{k\to 0}[ \cot\delta(k)-(2/\pi)\ln (k/2)+\gamma]$ exists. In (\ref{eq:cotdelta-3}), both integrals are of the type $\int_0^\infty F(k,r)\dd r$ where $\lim_{k\to 0} F(k,r)$ exists for fixed $r$ and $| F(k,r)|< B(r)$ with $\int_0^\infty B(r)\dd r<\infty$. Under this condition
\bel{eq:int-F}
\lim_{k\to 0}\int_0^\infty F(k,r)\dd r=\int_0^\infty F(0,r)\dd r~.
\ee
Hence we get
\bel{eq:lim-cot}
\lim_{k\to 0}\left\{ \cot\delta(k)-\frac{2}{\pi}\left[\ln\left(\frac{k}{2}\right)+\gamma\right]\right\}=
\frac{2}{\pi}\,\frac{-1+g X_2}{g X_1}=\frac{2}{\pi}\, \ln a~,
\ee
where $X_1$ and $X_2$ are given by Eqs.~(\ref{eq:defX1}) and (\ref{eq:defX2}), and $\ln a$ defined in Eq.~(\ref{eq:lna}).

Thus we finally have
\bel{eq:lim-cot-a}
\lim_{k\to 0}\left\{ \cot\delta(k)-\frac{2}{\pi}\left[\ln\left(\frac{k a}{2}\right) +\gamma\right]\right\}=0~.
\ee
As shown before, this limit is uniform for small $k$. Hence as $k\to 0$, $k>0$, we have
\bel{eq:lim-cot-b}
 \cot\delta(k)-\frac{2}{\pi}\left[\ln\left(\frac{k a}{2}\right) +\gamma\right]=\mathit{o}(1)~.
 \ee
 This immediately leads to 
 \bel{eq:lim-tan}
 \tan\delta(k)-\fracd{\pi/2}
 {\ln(ka/2) +\gamma}=\frac{\mathit{o}(1)}{| \ln k|^2}~.
 \ee
Thus under our general conditions on the potential $V(r)$, the length $a$ completely determines the $\mathit{O}(|\ln k|^{-2})$ contribution to $\delta(k)$.

Notice that our conditions (\ref{eq:intV1}) and (\ref{eq:intV2}), which can be summarized as
\bel{eq:intV12}
\int_0^\infty \left(1+|\ln r| + (\ln^+ r)^2\right) |V(r)|\,r\dd r <\infty~,
\ee
are in fact \emph{weaker} than those used in \cite{Chadan:1998qm}.

We now turn to the case $X_1(g)=0$. Notice first that if $X_1=0$, $1-g X_2\neq0$ because otherwise, from (\ref{eq:Volt1-0a}) we would get $u_0(r)/\sqrt{r}\to 0$ for $r\to\infty$, which implies $u_0(r)\equiv 0$. So for $X_1=0$, $u_0\propto \sqrt{r}$, which corresponds to a zero-energy bound state and 
\bel{eq:cor-X1z}
\cot\delta(k)-\frac{2}{\pi}\left(\ln(k/2)+\gamma\right)\to\infty\quad\text{for}\quad k\to 0~.
\ee

If $g_c$ is a value such that $X_1(g_c)=0$, $X_1$ has a \emph{single} zero at $g=g_c$, because the successive zeros of $X_1$ correspond to an increasing number of nodes for the zero-energy wave-function (this would not be true for the non-central case!). So, since $1-g_c X_2(g_c)\neq 0$, we have 
\bel{eq:sp-ca}
\begin{aligned}
\text{either}\quad&\begin{cases} a\to\infty\quad &\text{for}\quad g\to g_c\quad g<g_c~,\\
a\to0\ &\text{for}\quad g\to g_c\quad g>g_c~,\end{cases}\\
\text{or}\quad&\begin{cases} a\to 0\quad &\text{for}\quad g\to g_c\quad g<g_c~,\\
a\to\infty\quad &\text{for}\quad g\to g_c\quad g>g_c~.\end{cases}
\end{aligned}
\ee
\section{The weak coupling case}\label{se:2g}
In this section, we shall reconcile the apparent contradiction between the sign of $\tan\delta(k)$ as given by first order perturbation theory for small $g$, real $k>0$, and on the other hand the universal sign for $\tan\delta(k)$ given by our results which for small $k$ are independent of $V(r)$.

We will also explore further the behavior of the scattering length, $a$, as $g\to 0$. This turns out to be dependent on whether $g\int_0^\infty r\,V(r)\dd r$ is positive or negative.

From Eq.~(\ref{eq:cotdelta-1}), we have
\bel{eq:tandelta-1}
\tan\delta(k)=\fracd%
{\frac{\pi}{2}\,g \int_0^r J_0(kr')\,V(r')\,u(k,r')\,\sqrt{r'}\dd r'}
{\frac{\pi}{2}\,g \int_0^r Y_0(kr')\,V(r')\,u(k,r')\,\sqrt{r'}\dd r' - 1}~,
\ee
The power series in $g$ for $u(k,r)$ is  absolutely convergent. Thus for any real $k>0$, $\tan\delta(k)$ is given by the ratio of two entire functions in $g$. The perturbation series for $\tan\delta(k)$ has a radius of  convergence determined by the smallest zero, $|g_0|>0$, of the denominator.

Thus for small $|g|$, we have 
\bel{eq:tand-g}
\tan\delta(k)=-\frac{\pi}{2} \, g\, \int_0^\infty r \left[J_0(kr)\right]^2\,V(r)\,\dd r+\mathit{O}(g^2)~,
\ee
since $u(k,r)=\sqrt{r}\,J_0(kr)+\mathit{O}(g)$.

On the other hand, we have from Eq.~(\ref{eq:lim-tan})
 \bel{eq:lim-tan-bis}
 \tan\delta(k)-\fracd{\pi/2}{\ln(ka/2) +\gamma}=\frac{\mathit{o}(1)}{| \ln k|^2}~.
 \ee
The sign of $\tan\delta(k)$ in Eq.~(\ref{eq:tand-g}) changes when one goes from a repulsive $g V(r)$  to an attractive force. To reconcile this with Eq.~(\ref{eq:lim-tan}) or (\ref{eq:lim-tan-bis}), we need to study further the dependence of $a$ on $g$ near $g=0$.

From Eq.~(\ref{eq:lna}), we have 
\bel{eq:lna-1}
\ln a=\fracd{g \int_0^\infty \sqrt{r}\,\ln r\, V(r)\, u_0(r)\dd r -1}
{g \int_0^\infty \sqrt{r}\,V(r)\, u_0(r)\dd r }~,
\ee
where $u_0(r)$ satisfies the Volterra equation (\ref{eq:Volt1-0}), which to first order in $g$ reads
\bel{eq:Volt1-0-1}
u_0(r)=\sqrt{r}+g\,\sqrt{r}\,\int_0^r r' \,\ln\frac{r}{r'}\,V(r') u_0(r') \dd r' +\mathit{O}(g^2)~.
\ee

Substituting this result in (\ref{eq:lna-1}), we get for $a$
\bel{eq:a-2}
a=\exp\left[ \fracd{-1/g+   \int\limits_0^\infty r\,\ln r\,V(r)\dd r}{ \int_0^\infty r\,V(r)\dd r} +
\fracd{\int\limits_0^\infty r\,V(r)\dd r\int\limits_0^r r'\,V(r')\,\ln\frac{r}{r'}\dd r'}{\left(\int_0^\infty r\,V(r)\dd r\right)^2}
+\mathit{O}(g)\right]~,
\ee
Hence if $\int_0^\infty r\,V(r)\dd r>0$, $a(g)\to 0$ as $g\to 0$. However, if $\int_0^\infty r\,V(r)\dd r<0$, then as $g\to 0$ from $g>0$, $a\to\infty$. 

\begin{figure}[!htb]
\begin{minipage}{.5\textwidth}
\centerline{\includegraphics[width=.9\textwidth]{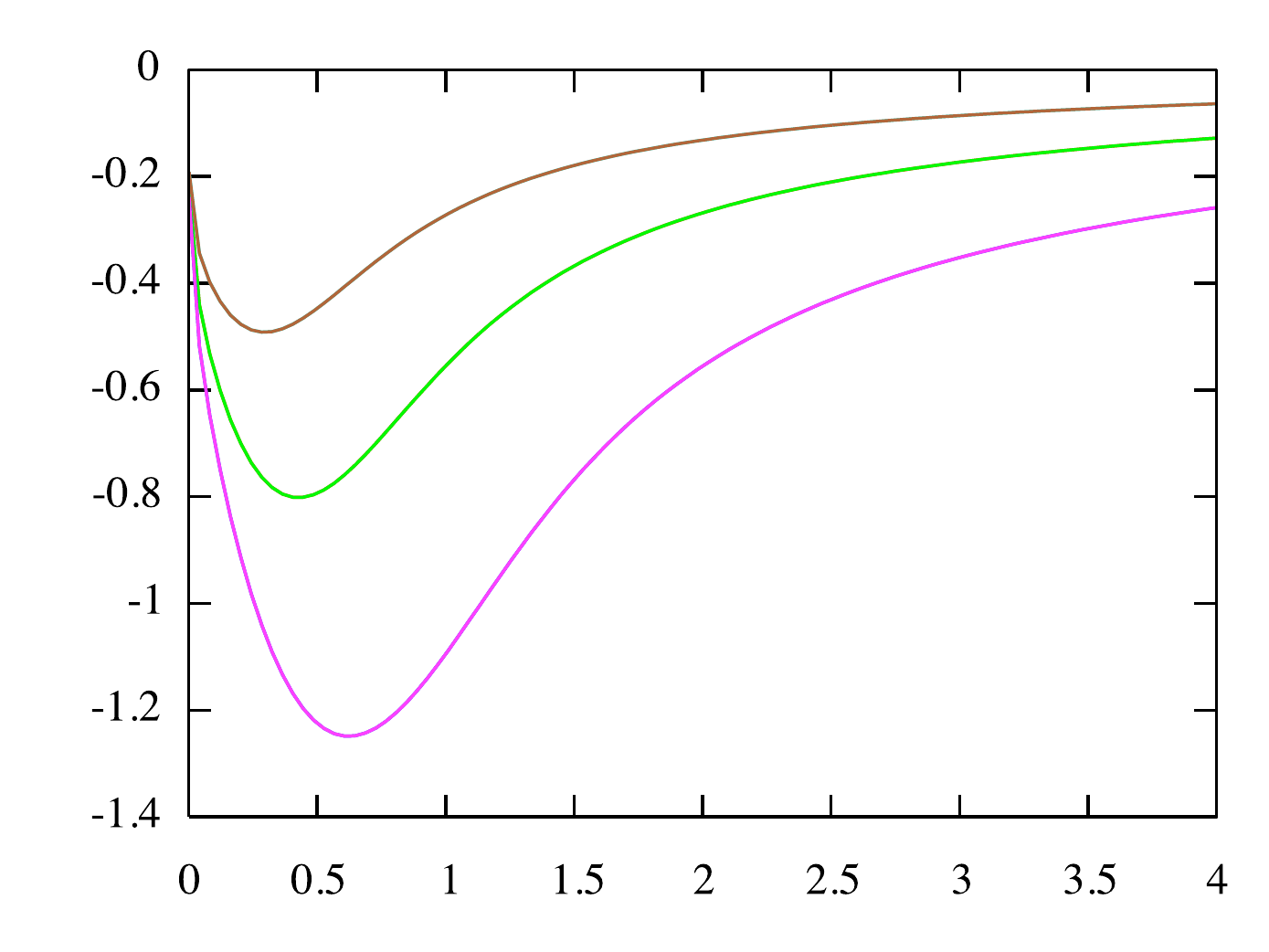}}
\end{minipage}
\begin{minipage}{.49\textwidth}
\caption{\label{fig:rep1}$\delta(k)$ for $V=g\exp(-r)$, and $g=1/2, \, 1,\, 2$. Each phase-shift is computed by two methods, but the two curves cannot be distinguished.}
\end{minipage}
\end{figure}

In the first case, $\int_0^\infty r\,V(r)\dd r>0$, there is no bound state, $\tan\delta <0$, the scattering length $a$ is small, and for small $k$, $\delta(k) \simeq (\pi/2)/\ln ka$.
In Fig.~\ref{fig:rep1}, $\delta(k)$ is drawn for the potential $g \exp(-r)$, and several values of $g$.
It is clearly observed that, while for $k\sim 2-4$, $\delta(k)$ is nearly proportional to $g$, for small $k$ there is an universal behavior.

In this figure, the phase-shift is computed by two methods.  First, the radial equation (\ref{eq:radialeq}) is solved numerally and $u(k,r)$ is matched into the Bessel functions, as per Eq.~(\ref{eq:ukasy}). For cross-checking, the radial equation is transformed into a non-linear, first order differential equation in $r$ for the phase function $\delta(k,r)$ of Calogero \cite{Calogero:224676}, and then $\delta(k)=\lim_{r\to\infty}\delta(k,r)$.

\begin{figure}[!htb]

\centerline{%
\includegraphics[width=.45\textwidth]{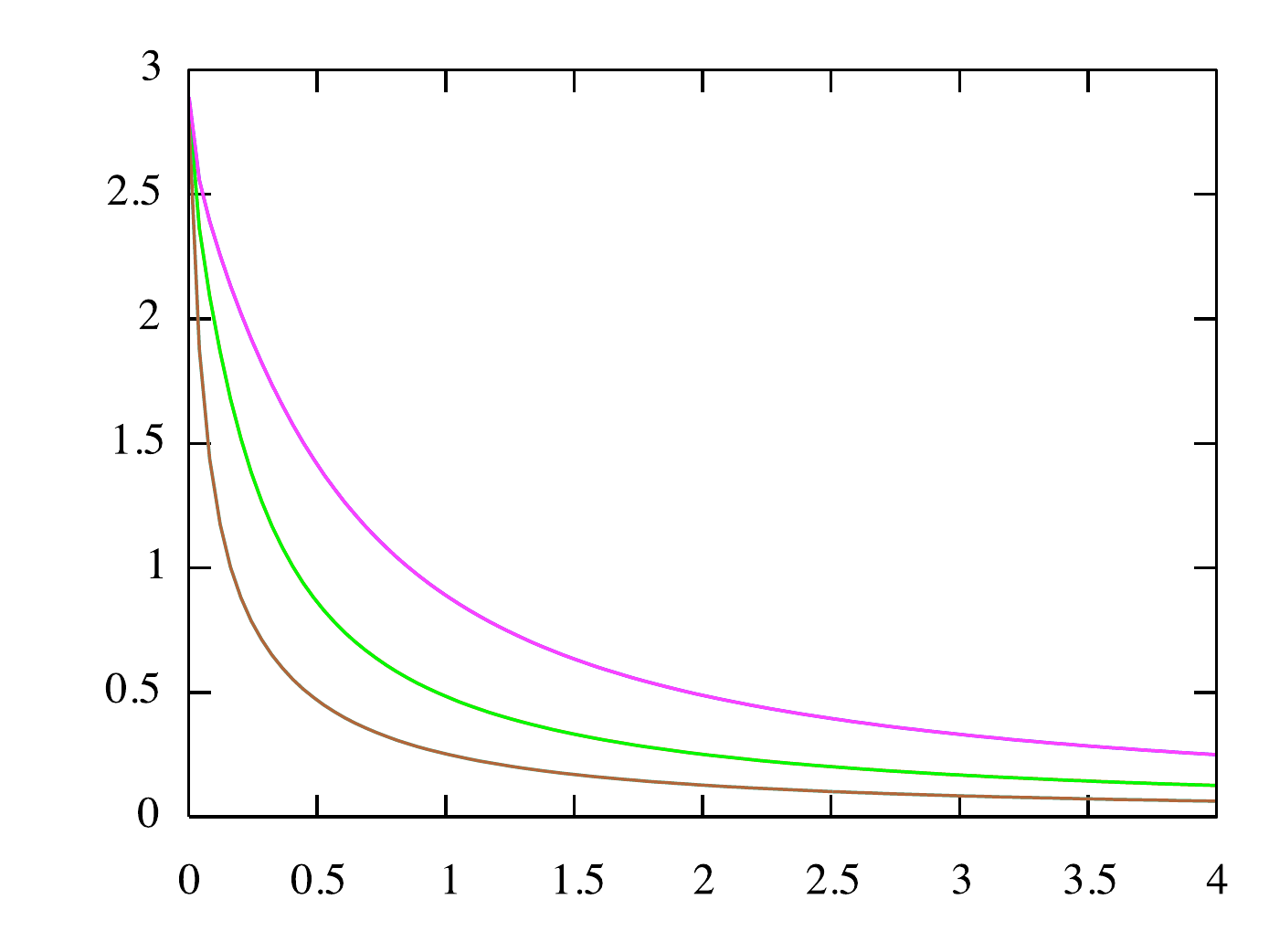}
\hfil
\includegraphics[width=.45\textwidth]{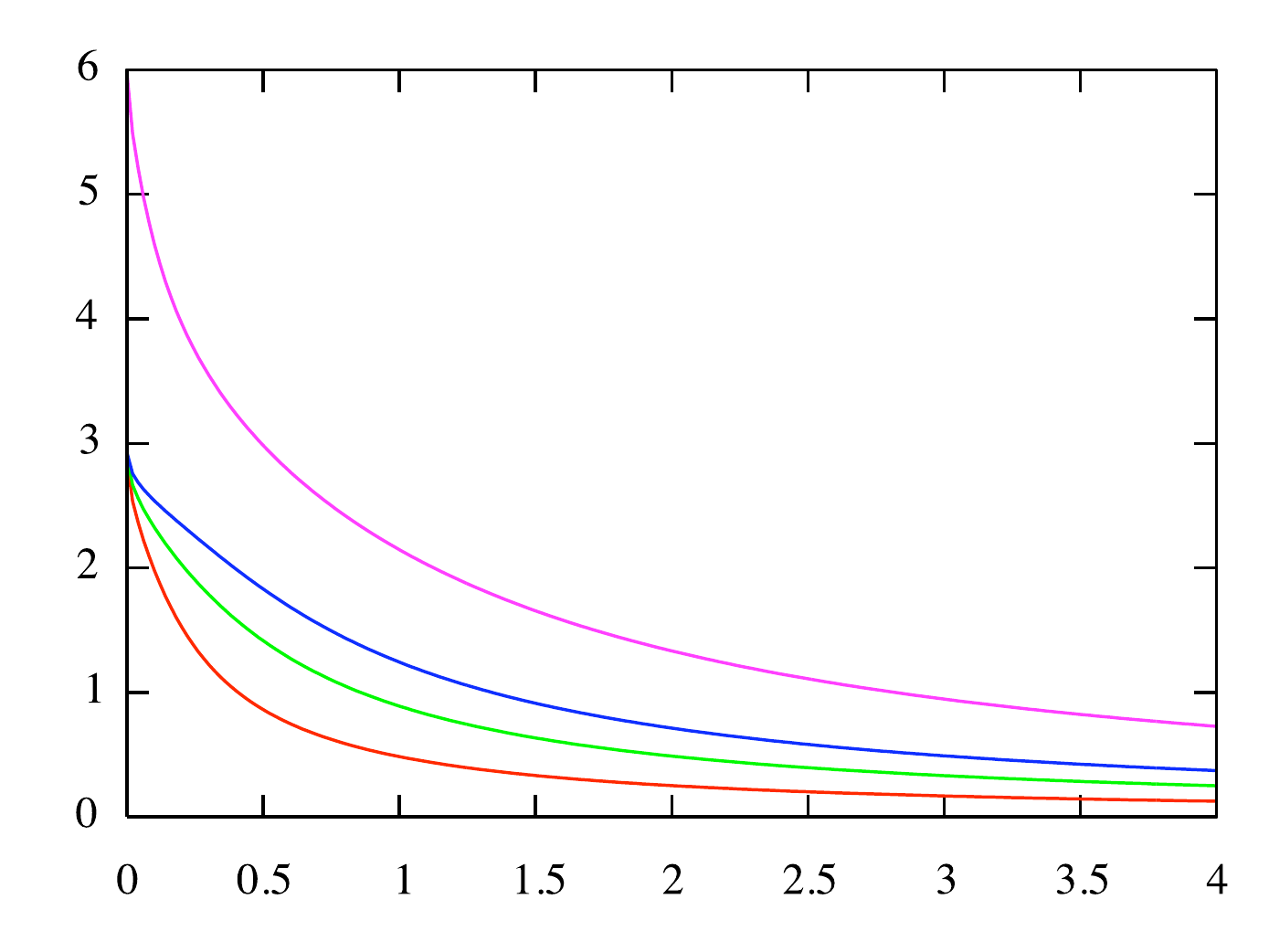}
}
\caption{\label{fig:att1}$\delta(k)$ for $V=-g\exp(-r)$, and $g=1/2, \, 1,\, 2$ (left) and 
$g=1,\, 2,\,3$ and $6$ (right).}
\end{figure}

For the second case $\int_0^\infty r\,V(r)\dd r<0$, we have at least one bound state for $g>0$, even if $g$ is small. However, the scattering length $a$ is large for small $g$, and the asymptotic result $\delta(k)\sim1/\ln ka$ will only be reached for very small $k$, $k\ll 1/a$. The phase-shift behaves as shown  in Fig.~\ref{fig:att1}.  For large $k$, $\tan\delta(k)>0$, in agreement with Eq.~(\ref{eq:tand-g}). An illustration is given in Fig.~\ref{fig:att1}, for the potential $V=-g\exp(-r)$ and several values of the coupling $g$.

Finally, one should note that Figs.~\ref{fig:rep1} and \ref{fig:att1} illustrate the Levinson theorem, whose validity in 3D was established by one of us, A.M. (and many others!), many years ago 
\cite{martinlev2d,PhysRevA.56.1938,PhysRevLett.56.900}, and this particular proof can be extended to 2D.
The convention  adopted in Fig.~\ref{fig:att1} is that $\delta(k)\to 0$ as $k\to\infty$. Hence $\delta(0)= n \pi$,  according to the number $n$ of bound states.
%
\section{The effective range in 2D}\label{se:2R}
Since we have been able to introduce the scattering length in 2D in analogy to 3D, it is natural to ask oneself if one can also define and effective range in 2D. In fact this has already been done by a group of Dutch physicists \cite{0305-4470-17-3-020}. However, what we want to do here is to find the conditions under which the effective range exists and what happens if it is infinite. We shall see that the condition (\ref{eq:intV12}) is not sufficient and that, crudely speaking, we need a potential decreasing faster at infinity.

The method we shall use is, at least at the beginning, a carbon copy of the method used by Blatt and Jackson \cite{PhysRev.76.18}, Blatt and Weisskopf \cite{Blatt:628052} and other textbooks 
\cite{Joachain:104904}.

From now on, we shall use a \emph{different} normalization for the wave function. First we define the free zero-energy wave function $v_0(r)$ which coincides with $u_0(r)$ for $r\to\infty$ as
\bel{eq:v0n}
v_0(r)=\sqrt{r}\,\ln\left(\frac{r}{a}\right)~,
\ee
and the free wave function $v(k,r)$ which coincides with the exact wave function at infinity will be normalized in such a way that 
\bel{eq:vn}
v(k,r)\to v(0,r)=v_0(r)~,
\ee
for any finite $r$, as $k\to 0$, or more generally for any $r(k)$ such that $k\,r(k)\to 0$ for $k\to 0$. 
One such normalization is
\be \label{eq:freev}
v(r)= -\frac{\pi}{2}\sqrt{r} \left[ \cot\delta\,J_0(kr) - Y_0(kr)\right]~.
\ee
On can check that in the limit $k\to 0$, the coefficient of $\sqrt{r}\,\ln r$ is correct, and that $\cot\delta$ disappears   because of Eq.~(\ref{eq:lim-cot-b}). 

The standard procedure \cite{Blatt:628052} is to combine the exact and free equations for $k=0$ and for $k\neq 0$ to get
 \be 
\lim_{r\to 0}[ v\, v'_0-v' v_0]=k^2\int_0^\infty(u\,u_0-v\,v_0)\dd r~,
 \ee
 and substituting (\ref{eq:v0n}) and (\ref{eq:freev}),
 \be \label{eq:ex2D}
 \cot\delta-\frac{2}{\pi}\left[\ln\left(\frac{k a}{2}\right)+\gamma\right]=k^2 \int_0^\infty(v v_0-u u_0)\dd r~.
 \ee
 This expression is  \emph{exact}. The crucial step is to say
  \be \label{eq:app2D}
\lim_{k\to 0}\frac{1}{k^2}\left\{ \cot\delta-\frac{2}{\pi}\left[\ln\left(\frac{k a}{2}\right)+\gamma\right]\right\}= \int_0^\infty(v_0^2-u_0^2)\dd r~,
 \ee
 where we use the fact that since $v$ approaches $v_0$ and since the normalization of $u$ is fixed by that of $v$, $u$ approaches $u_0$. However, there is absolutely no guarantee that the integral in the R.H.S.\ of (\ref{eq:app2D}) converges.
 
 Since $u_0$ approaches $v_0$, it is sufficient to study $(u_0-v_0) v_0$. Now, we use again a Volterra equation, but contrary to what was done in Sec.~\ref{se:a-2d}, we start from infinity, i.e., 
\be \label{eq:uz-vz}
u_0=v_0+g\,\sqrt{r}\int_r^\infty\ln\left(\frac{r'}{r}\right)\sqrt{r'}\,V(r')\,u_0(r')\dd r'~,
\ee
Since we only want to have the behavior of $u_0-v_0$ for large $r$, it is sufficient to take the first term of the perturbative expansion of (\ref{eq:uz-vz}), i.e.,
\bel{eq:uz-vz-a} 
u_0-v_0\simeq g\,\sqrt{r}\int_r^\infty\ln\left(\frac{r'}{r}\right)\sqrt{r'}\,V(r')\,v_0(r')\dd r'~.
\ee
The integral in (\ref{eq:app2D}) will converge  if
\bel{eq:Vr3ln2} 
\int_0^\infty v_0(r)\, \sqrt{r}\dd r\int_r^\infty \ln\left(\frac{r'}{r}\right)\sqrt{r'}\,V(r')\, v_0(r')\dd r'<\infty~,
\ee
and substituting (\ref{eq:v0n}) we get in the end
\be \label{eq:condr3l2}
\int_0^\infty V(r)\, r^3\left[\ln\left(\frac{r}{a}\right)\right]^2 \dd r<\infty~.
\ee

This is a \emph{sufficient} condition for the existence of the effective range, defined by (\ref{eq:app2D}). However, if we restrict ourselves to potentials with a \emph{definite} sign for large $r$, this condition is also \emph{necessary}. If  (\ref{eq:Vr3ln2}) diverges, the effective range is infinite. Hence (\ref{eq:Vr3ln2}) can be regarded as the best possible condition.

Now, can we say something more when (\ref{eq:Vr3ln2}) diverges? We have investigated the case where 
\bel{eq:r-nu}
V(r)=g\,r^{-\nu} \quad\text{for}\quad r>R\quad \text{with} \quad 2<\nu\le4~.
\ee
Notice that $\nu<2$ is incompatible with (\ref{eq:intV12}). The strategy is to write the R.H.S.\ of (\ref{eq:ex2D}) as
\bel{eq:uv-split}
k^2 \int_0^\infty \left[(u-v) u_0+(u_0-v_0) v \right]\dd r~,
\ee
and use not only (\ref{eq:uz-vz}) but also the Volterra equation for $u$
\be \label{eq:u-v}
u=v+g\,\sqrt{r}\int_{r}^{\infty}\left[J_0(kr)Y_0(kr')-J_0(kr')Y_0(kr)\right]\sqrt{r'}\,V(r')\, u(r')\dd r'~,
\ee
and, since we are only interested in large $r$, we keep only
\be\label{eq:u-v-a}
u-v\simeq g\,\sqrt{r}\int_{r}^{\infty}\left(J_0(kr)Y_0(kr')-J_0(kr')Y_0(kr)\right)\sqrt{r'}\,V(r')\, v(r')\dd r'~,
\ee

Now comes the tedious, but straightforward task of estimating (\ref{eq:uv-split}) by inserting 
(\ref{eq:uz-vz-a}) and (\ref{eq:u-v-a}), and the known expressions for $v$ and $v_0$.  Details are given in Appendix \ref{sec:appb}. The final answer is
\begin{multline}\label{eq:r-nu-2d}
\cot\delta(k)-\frac{2}{\pi}\left[\ln\left(\frac{k a}{2}\right)+\gamma\right]\\
\simeq 
-\frac{2}{\pi} g\,k^{\nu-2}\left[\ln (ka)\right]^2\,\frac{1}{\nu-2}\left(\frac{1}{2}\right)^{\nu-3}\,
\fracd{\Gamma(\nu-2)\Gamma(2-\nu/2)}{\Gamma^2(\nu/2) \Gamma(\nu/2-1)}~,
\end{multline}
if $V=g\,r^{-\nu}$ for $r>R$.
%
%
\section{Effective range expansion in 3D}\label{se:3R}
Even though the notion of effective range in 3D is well known (see, e.g., \cite{PhysRev.76.18,Blatt:628052,Joachain:104904} and  references there to other pioneering contributions), we want to come back on the subject, because we believe that, at least to the best of our knowledge, its validity has not been investigated systematically. 
Mott and Massey \cite{Mott:104647} explain that if the potential decreases at large $r$ as
$r^{-s}$, the effective range will exist only for $s>5$%
\,\footnote{We thank the referee for a clarification on this point.}%
. We shall give a somewhat broader sufficient condition, which could be the equivalent of (\ref{eq:condr3l2}) in 3D.
 K.~Chadan has given to us a series of references on cases where the effective range formula is \emph{not} valid \cite{handelsman68}, namely for a potential behaving like $g\,r^{-\nu}$ at large $r$, with $3<\nu<5$, one has
\be\label{eq:ere3D-nu}
k\,\cot\delta +\frac{1}{a}\sim g\,C(\nu)\,k^{\nu-5}~,
\ee
where $C(\nu)$ is known. Again, this is analogous to Eq.~(\ref{eq:r-nu-2d}) in 2D.

Now, the effective-range formula in 3D is
\be\label{eq:ere3D}
k\,\cot\delta+\frac{1}{a}\simeq \frac{1}{2}\, r_0\, k^2~,
\ee
for $k\to 0$, where the effective range is given by
\bel{eq:rz3D}
r_0=2\int_0^\infty(v_0^2-u_0^2)\dd r~,
\ee
 (the factor $1/2$ is such that $r_0$ is  about the radius in the case of square-well potential, see, e.g., \cite{Joachain:104904}). Here $v_0$ is the free reduced wave function at zero energy, with a normalization 
 \bel{eq:v0n3d}
 v_0=1-\frac{r}{a}~,
 \ee
 and $u_0$ is the exact solution normalized in such way that $u_0-v_0\to 0$ as $r\to\infty$. So $u_0$ satisfies the Volterra equation
\bel{eq:3dVz1}
u_0(r)=v_0(r)+g\int_r^\infty(r'-r)\,V(r') u_0(r')\dd r'~.
\ee
The asymptotic behavior of $u_0-v_0$ is therefore given by the lowest order iteration of (\ref{eq:3dVz1})
\bel{eq:3du0mv0}
u_0-v_0\simeq \int_r^\infty (r'-r)\,V(r')\,(1-r'/a)\dd r'~,
\ee
and the convergence of  (\ref{eq:rz3D}) will depend on the convergence of the integral
\bel{eq:int-ra-3D}
\int_R^\infty (1-r/a)\dd r\int_r^\infty (r'-r)\,V(r')\,(1-r'/a)\dd r'~.
\ee
The conclusion is that the effective-range formula will hold if 
\bel{eq:cond-3D-rz}
\int_0^\infty  r^4 \left| V(r)\right| \dd r<\infty~.
\ee
In addition, to get the existence of  a scattering length, we should impose
\bel{eq:cond-a-3D}
\begin{aligned}
&\int_0^\infty r\phantom{^2} \left| V(r)\right| \dd r<\infty~,\\
&\int_0^\infty r^2 \left| V(r)\right| \dd r<\infty~.
 \end{aligned}
 \ee

If $V$ has constant sign for large $r$, the condition (\ref{eq:cond-3D-rz}) is not only sufficient, but also \emph{necessary}.  Then, if (\ref{eq:cond-3D-rz}) diverges, 
\bel{eq:div-3D}
\frac{k\,\cot\delta(k)+1/a}{k^2}\to\infty~,
\ee
for $k\to 0$. In particular, the existence of an effective range cannot be summarized as whether or not $r^5V(r)\to 0$ at large $r$. For instance
\begin{itemize}
\item
For the potential
\bel{eq:counter1}
V(r)=\frac{\Theta(r-R)}{r^5\,\ln^\alpha(r+R)}~,
\ee
such that $r^5\,V(r)\to 0$, the effective-range formula is  valid only if $\alpha>1$.
\item
For the potential
\bel{eq:counter2}
V(r)=\exp\left[-r^{12}\,\sin^2 r\right]~,
\ee
such that $V(r)=1$ for  $r=n\pi$, $n\in\mathbbm{N}$, the effective range formula is \emph{valid}, even though $V(r)$ does not decreases faster than $r^{-5}$. In this case, it is better to think of (\ref{eq:cond-3D-rz}) as a Lebesgue integral.
\end{itemize}

A final remark: $r_0$ is generally believed to be positive, because at short distances, $v_0$ is much larger than $u_0$. In fact, this is not necessarily true. Take the potential
\bel{eq:counter3}
V(r)=-2\,\frac{\delta(r-R)}{R} -\frac{3\,R\,\delta(r-3R-D)}{D\,(R+D)}~.
\ee
It is easy to see that it has a scattering length $a=3R$ for any $D>0$, and that the effective range goes to $-\infty$ for $D\to\infty$, while it is positive for $D$ close to $0$.
\section{Outlook}
In this paper, we have presented a rigorous definition of the scattering length in two dimensions, both from the zero of the zero-energy solution and from the low-energy behavior of the $S$-wave phase-shift, and the equivalence between the two definitions has been demonstrated. The effective-range expansion is also given for short-range potentials, and conditions have been written for the existence of a scattering length and  of an effective range. The effective-range expansion is also generalized for classes of power-law potentials with a longer range. 

The main result is that for short-range potentials, the S-wave phase-shift $\delta(k)$ behaves at low energy such that
\bel{eq:sum}
\cot\delta=\frac{2}{\pi}\left[\ln(ka/2)+\gamma\right]+ \frac{1}{2}\,r_0\,k^2+\cdots~,
\qquad
r_0=2\,\int_0^\infty(v_0^2-u_0^2)\,\d r~,
\ee
where the zero-energy reduced wave-function $u_0(r)$ has an asymtotic form $u_0(r)\to v_0(r)$, with $v_0(r)$ being the free, zero-energy solution $v_0(r)=\sqrt{r}\,\ln (r/a)$.

This investigation gave us the opportunity to revisit the same questions in three dimensions, and to clarify and expand the existing results. An alternative derivation of our conditions (displayed in the first preprint version of this article)  have been  proposed recently by Chadan \cite{Chadan:1151902}.

We believe that it is very unlikely that the existence of the scattering length in two or three dimension depends on the rotational symmetry of the potential $V$. In other words, under very general conditions on the potential $V(\vec{r})$, without rotational symmetry, the scattering can be expected to be well defined. It is our intend to study this problem.

\begin{acknowledgments}
We thank Khosrow Chadan for informative discussions on previous work on the subject. One of us (T.T.W.) would like to thank the hospitality provided to him by the CERN theory division.
\end{acknowledgments}

\appendix
\section{Bound on $R(z)$}\label{se:Appa}
 The function $R(z)$ defined by Eqs.~(\ref{eq:R}) has the following properties.
\begin{enumerate}
\item
From \cite{Stegun:102084},
\be\label{eq:Rz}
R(z)=\frac{8}{\pi^2}\int_0^{\pi/2}\cos (z\cos\theta)\ln(2\sin\theta)\dd\theta~,
\ee
\item
\be
R(0)=0~,\quad \hbox{because}\quad \int_0^{\pi/2} \ln (2\sin\theta)\dd\theta=0~.
\ee
Proof:
\be\begin{aligned}
 \int_0^{\pi/2} \ln (2\sin\theta)\dd\theta&= \int_0^{\pi/2} \ln (2\cos\theta)\dd\theta=\frac{1}{2} \int_0^{\pi/2} \ln (2\sin 2\theta)\dd\theta\\
 &=\frac{1}{4} \int_0^{\pi} \ln (2\sin\phi)\dd\phi=\frac{1}{2} \int_0^{\pi/2} \ln(2 \sin(\phi)\dd\phi~.
\end{aligned}\ee
which would lead to a contradiction unless it vanishes.
\item
\be
\left| R(z)\right| < \frac{8}{3\pi}~.
\ee
Proof: 
\be
\left| R(z)\right|< \frac{8}{\pi^2}\int_0^{\pi/2}\left|\ln(2\sin\theta)\right|\dd\theta=
\frac{16}{\pi^2}\int_0^{\pi/6}\left|\ln(2\sin\theta)\right|\dd\theta~,
\ee
But, due to the convexity of $\sin \theta$ for $\theta\in[0,\pi/6]$, $2\sin\theta>\theta/(\pi/6)$ in this interval, and the above integral is less than $\pi/6$, and hence
\be
|R(z)|< \frac{8}{3\pi}~.
\ee
On the other hand, introducing
\be
I(\theta)=\int\limits_0^\theta \ln(2\sin\theta)\dd\theta~,
\quad I(0)=I(\pi/2)=0~,\quad I(\theta)<0\ \hbox{if}\ \theta\in]0,\pi/2[~,
\ee
we get by integrating by parts
\be
R(z)=\frac{8}{\pi^2}\left[\left[\cos(z \cos\theta)\,I(\theta)\right]_0^{\pi/2}-
z\int_0^{\pi/2}\sin(z\cos\theta)\,\sin\theta\,I(\theta)\dd\theta\right]~,
\ee
so
\be
|R(z)|<-\frac{8 z^2}{\pi^2}\int_0^{\pi/2}\sin\theta\,\cos\theta\,I(\theta)\dd\theta~,
\ee
but by integrating again by parts,
\be
\int_0^{\pi/2}\sin\theta\,\cos\theta\,I(\theta)\dd\theta=-\frac{1}{4}\int_0^{\pi/2}\cos 2\theta \,\ln(2\sin\theta)\dd\theta=
\ee
so
\be |R(z)|<\frac{z^2}{2\pi}~, \quad\hbox{and}\quad\lim_{ r\to 0}\frac{|R(z)|}{z^2}=\frac{1}{2\pi}~.\ee
Also
\be\begin{aligned}
-J_0(z)+1&=\frac{1}{\pi}\int_0^\pi\left[1-cos(z\cos\theta)\right]\mathrm{d}\theta=
\frac{2}{\pi}\int_0^\pi\sin^2[(z\cos\theta)/2]\dd\theta\\
&<\frac{2}{\pi}\int_0^\pi[(z\cos\theta)/2]^2\dd\theta=\frac{z^2}{4}~.
\end{aligned}\ee
\end{enumerate}
\section{Details about the effective range in 2D}\label{sec:appb}
 We want to estimate (\ref{eq:uv-split}) in the case where $\int_0^\infty(v_0^2-u_0^2)\dd r$ diverges. Since  (\ref{eq:uv-split}) is dominated by the large $r$ behavior, it is sufficient to use the asymptotic form (\ref{eq:uz-vz-a}) of $u_0-v_0$ and (\ref{eq:u-v-a}) fo $u-v$. So we have
 \bel{eq:split-B}
 \int_0^\infty\left[(v_0\,(u-v)+v\,(u_0-v_0)\right]\simeq X+Y~,
 \ee
 where
 \begin{multline}\label{eq:B-X}
 X=g\frac{\pi}{2}\int_R^\infty r\,\ln\left(\frac{r}{a}\right)\d r \int_r^\infty
 \left[J_0(kr) Y_0(kr')-J_0(kr') Y_0(kr)\right] r'\\
 \times
 V(r') \left[\cot\delta\,J_0(kr')-Y_0(kr')\right]\d r'~,
 \end{multline}
 and
  \bel{eq:B-Y}
Y=g\int_R^\infty  r  \left[\cot\delta\,J_0(kr)-Y_0(kr)\right]\d r
\int_0^\infty  \ln\left(\frac{r'}{r}\right)V(r')\,r'\,\ln\left(\frac{r'}{a}\right)\dd r'
 \ee
or, by exchanging the order of integration,
\begin{multline}\label{eq:B-Xa}
X=g\frac{\pi}{2}\int_R^\infty r \,V(r)\left[\cot\delta\,J_0(kr)-Y_0(kr)\right] \d r\\ \times
\int_R^r r'\,\ln\left(\frac{r'}{a}\right) \left[J_0(kr) Y_0(kr')-J_0(kr') Y_0(kr)\right]\d r'~,
\end{multline}
and
  \bel{eq:B-Y-a}
Y=g \int_R^\infty r\,  \ln\left(\frac{r}{a}\right)V(r)\dd r
\int_R^r r'\, \ln\left(\frac{r'}{r}\right)  \left[\cot\delta\,J_0(kr')-Y_0(kr')\right] \d r'~.
 \ee
If we restrict ourselves to $V(r)=r^{-\nu}$ for $r>R$, we can use scaling, taking the variable $z=k\,r$, and get
\begin{multline}\label{eq:B-Xb}
X=g\frac{\pi}{2}\, k^{\nu-4} \int_{kR}^\infty \d z  \,z^{1-\nu}\left[\cot\delta\,J_0(z)-Y_0(z)\right]\\ \times
\int_{kR}^z z'\,\ln\left(\frac{z'}{ka}\right) \left[J_0(z) Y_0(z')-J_0(z') Y_0(z)\right]\d z'~,
\end{multline}
and
  \bel{eq:B-Y-b}
Y=g k^{\nu-4} \int_{kR}^\infty z^{1-\nu}\,  \ln\left(\frac{z}{ka}\right) \dd z
\int_{kR}^z z'\, \ln\left(\frac{z'}{z}\right)  \left[\cot\delta\,J_0(z')-Y_0(z')\right] \d z'~.
 \ee
Unless we have convergence problems at the origin, we can replace $kR$ by zero for $k\to 0$. Retaining only the \emph{dominant} terms, we get
\begin{multline}
\cot\delta-\frac{2}{\pi}\left[ \ln \left(\frac{k a}{2}\right) + \gamma\right] \simeq
-k^{\nu -2}\left[\ln\left(\frac{k a}{2}\right)\right]^2 \left[
 \int_0^\infty \d z \, z^{1-\nu}\,\frac{2}{\pi} \int_0^z \d z'\, z' \,\ln\left(\frac{z}{z'}\right) J_0(z')\right.\\
\left. {}+\int_0^\infty \d z \, z^{1-\nu}\,J_0(z) \int_0^z \d z'\, z' \left[J_0(z') Y_0(z) - J_0(z)Y_0(z')\right]\right]~,
\end{multline}

Now we have to evaluate certain integrals:
\begin{enumerate}
\item
From the Bessel differential equation
\be\begin{split} 
\int_0^z z' J_0(z')\dd z'&=-z\,J_0'(z)~,\\
\int_0^z z' Y_0(z')\dd z'&=-z\,Y_0'(z)+\frac{2}{\pi}
\end{split}\ee
\item
\be
\int_0^z x \ln\left(\frac{z}{x}\right) J_0(x)\dd x=1-J_0(z)~,
\ee
\item
\be\begin{split} 
\int_0^z z' J_0(z')\dd z'&=-z\,J_0'(z)~,\\
\int_0^z z' Y_0(z')\dd z'&=-z\,Y_0'(z)+\lim_{z\to 0} (z Y'_0(z)=-z\,Y_0'(z)+\frac{2}{\pi}
\end{split}\ee
 and using the Wronskian $Y'_0(z)J_0(z)-Y_0(z) J'_0(z)=2/(\pi z)$, we get
 \be
\int_0^z z'\dd z'\left[ J_0(z') Y_0(z)-J_0(z)Y_0(z')\right]
=\frac{2}{\pi} \left[1 -J_0(z)\right]~.
\ee
\end{enumerate}

Overall, we get
 \be
 \begin{split}
 \cot\delta-\frac{2}{\pi}\left[ \ln \left(\frac{k a}{2}\right) + \gamma\right] &\simeq
 g\, k^{\nu -2}\left[\ln\left(\frac{k a}{2}\right)\right]^2  \frac{2}{\pi} \int_0^\infty 
  z^{1-\nu} \left[1-[J_0(z)]^2\right]  \d z \\
&=  g\, k^{\nu -2}\left[\ln\left(\frac{k a}{2}\right)\right]^2  \frac{2}{\pi} \frac{2}{2-\nu} \int_0^\infty
J_0(z)\,J_1(z)\, z^{2-\nu}\dd z~,
\end{split}\ee
where the last integral can be evaluated as using a formula in \cite[p.\ 35]{Magnus:102686}, \cite[p.\ 715]{Gradshteyn:476809} and also \cite[Vol.~II, p.~52, Eq.~(30)]{Bateman:100230} where the conditions $2<\RE{\nu}<4$ are specified
\be
\int_0^\infty J_0(z)\,J_1(z)\, z^{2-\nu}\dd z=\frac{2}{\pi}\,\frac{1}{\nu-2} \left(\frac{1}{2}\right)^{\nu-3}
\frac{\Gamma(\nu-2)\Gamma(2-\nu/2)}{[\Gamma(\nu/2)]^2\,\Gamma(\nu/2-1)}~.
\ee


%

\end{document}